\documentclass[12pt]{article}
\input epsf.sty
\def\be{\begin{equation}}
\def\ee{\end{equation}}
\def\ba{\begin{align}}
\def\ea{\end{align}}

\def\p{\partial}

\def\ops[#1]{\p_{#1} e^{-2\phi}}

\def\eq[#1]{equation (\ref {eq:#1})}
\def\Eq[#1]{Equation (\ref {eq:#1})}
\def\e[#1]{\ref {eq:#1}}
\def\at[#1]{| _{#1}}

\let\oldpercent\%\renewcommand{\%}{\scalebox{0.85}{\oldpercent}}

\begin{document}

\baselineskip=18pt

\begin{center}
{\Large \bf{2pf in single-trace $T\bar T$ holography}}

\vspace{10mm}

\textit{ Amit Giveon}
\break

Racah Institute of Physics, The Hebrew University \\
Jerusalem, 91904, Israel

\end{center}


\vspace{10mm}

\begin{abstract}

The 2pf in single-trace $T\bar T$ holography was computed and analyzed for operators in the $w=0$ sector in~\cite{Asrat:2017tzd}.
Here we present its immediate generalization to non-zero winding $w$.
For long strings, the result is identical to the one obtained within the TsT/$T\bar T$ approach in~\cite{Cui:2023jrb}
(up to a factor of the reflection coefficient).

\end{abstract}
\vspace{10mm}

Single-trace $T\bar T$ holography~\cite{Giveon:2017nie} is obtained from the $AdS_3/CFT_2$ one by adding to the Lagrangian of the $CFT_2$ a certain dimension $(2,2)$ quasi-primary operator $D(x)$, $x\equiv(x^0,x^1)$, constructed in~\cite{Kutasov:1999xu}. This irrelevant deformation can be controlled, since in the dual bulk string theory on $AdS_3$, it corresponds to deforming the worldsheet theory, in which $AdS_3$ is described by the $SL(2,R)$ WZW model, by a current-current operator, $J^-\bar J^-$,~\cite{Giveon:2017nie}.

A useful description of the deformed worldsheet theory is obtained by a null gauging of the SCFT
\be\label{rads}
R^{1,1}\times AdS_3\times{\cal N}
\ee
in the directions
\be\label{null}
i\partial(y-t)+\alpha J^-~,\qquad i\bar\partial(y+t)\pm\alpha \bar J^-
\ee
where $(t,y)$ are the canonically normalized bosonic fields of the $R^{1,1}$ SCFT.
The resulting theory is obtained along the lines presented in~\cite{Giveon:2017myj,Asrat:2017tzd};
it is referred to as the superstring on
\be\label{m3}
{\cal M}_3\times{\cal N}
\ee
in~\cite{Giveon:2017nie,Giveon:2017myj,Asrat:2017tzd}.

Following~\cite{Asrat:2017tzd},
a large class of observables in the superstring on~(\ref{m3}) is given by
vertex operators in the (NS,NS) sector, which take the form (in the $(-1,-1)$ picture)
\be\label{o}
O^{w,j}(p)=\int d^2z e^{-\varphi-\bar\varphi}\Phi_{h}^{w,j}(p)e^{-i\alpha(p_0 t+p_1 y)}O_{\cal N}
\ee
Here $\varphi,\bar\varphi$ are worldsheet fields associated with the superconformal ghosts, that keep track of the picture, and
$\Phi_{h}^{w,j}(p)$ are vertex operators on $AdS_3$ in the winding $w\geq 0$ sectors,
obtained by spectral flow from operators in the representation $j$ of $SL(2,R)$;
they are the Fourier transforms of the operators $\Phi_h^{w,j}(x;z)$ in~\cite{Maldacena:2001km} to momentum space,
$p\equiv(p_0,p_1)$.~\footnote{In~\cite{Maldacena:2001km}, $h$ was denoted by $J$, and we consider operators with $\bar J=J\equiv h$.}

The operators in eq.~(\ref{o}) include in addition to the set of operators in the $w=0$ sector (in which case $h=j$),
considered in eq. (2.9) of~\cite{Asrat:2017tzd} (in which case the $AdS_3$ operator is $\Phi_h^{w=0,j=h}(x;z)\equiv\Phi_h(x;z)$),
also those with non-zero winding, $w>0$
(in which case the spacetime conformal weight $h$ of the $AdS_3$ operator $\Phi_h^{w>0,j}(x;z)$ and its $j$ are different quantum numbers).

When $w>0$, the mass-shell condition for the operators in eq.~(\ref{o}) is~\footnote{This is obtained using e.g. eq. (4.6) in~\cite{Giveon:2019fgr},
which follows from eqs. (3.4),(3.5) in~\cite{Maldacena:2001km};
we take $\bar\Delta_{\cal N}=\Delta_{\cal N}$, so that $\bar h=h$ below, as in~\cite{Asrat:2017tzd}.}
\be\label{onshell}
-{j(j-1)\over k}-w\left(h_{p^2}-{k\over 4}w\right)+{1\over 2}\lambda p^2+\Delta_{\cal N}={1\over 2}
\ee
where $\lambda=\pm\alpha^2$, with the $\pm$ amounting to axial and vector gauging in~(\ref{null}), respectively.
The subscript $p^2$ was added to $h$, to emphasize that it depends on $p^2$ via the on-shell condition.

In the $w=0$ sector, $j=h_{p^2}$, and the on-shell condition is
\be\label{wzero}
-{h_{p^2}(h_{p^2}-1)\over k}+{1\over 2}\lambda p^2+\Delta_{\cal N}={1\over 2}
\ee
This is eq. (3.5) in~\cite{Asrat:2017tzd}.
Equation~(\ref{onshell}) is the generalization of~(\ref{wzero}) to the $w>0$ sectors.

Denoting by $h$ the value of $h_{p^2}$ in the undeformed theory, $h\equiv\lim_{p\to 0}h_{p^2}$,
from eq.~(\ref{onshell}) one finds that, for $w\neq 0$,
\be\label{hp}
h_{p^2}=h+{1\over 2}{\lambda\over w} p^2
\ee

On the other hand, from eq.~(\ref{wzero}), one finds that, for $w=0$,
\be\label{hp0}
2h_{p^2}-1=\sqrt{(2h+1)^2+2\lambda kp^2}
\ee
instead. This is eq. (3.6) in~\cite{Asrat:2017tzd}.

Equation~(\ref{hp}) is the generalization of eq.~(\ref{hp0}) to the non-zero winding cases, $w\neq 0$.

The main result of this note is the following.

One finds that the 2pf that amounts to the worldsheet operators $O^{w>0,j}(p)$ in the target space is
\be\label{oo}
\langle \hat O^{w>0,j}(p)\hat O^{w>0,j}(-p)\rangle_{target}=R(w,j;h_{p^2})\gamma(1-2h_{p^2})\left(p^2\over 4\right)^{2h_{p^2}-1}
\ee
with the $h_{p^2}$ in~(\ref{hp}), and where
\be\label{a}
R(w,j;h_{p^2})=k\nu^{1-2j}{\gamma\left(j-{k\over 2}w+h_{p^2}\right)\gamma\left(j+{k\over 2}w-h_{p^2}\right)\over
\gamma\left({2j-1\over k}\right)\gamma(2j)}
\ee
is the bulk reflection coefficient in the superstring on ${\cal M}_3\times{\cal N}$,~(\ref{rads})--(\ref{m3}),~\footnote{For 
discrete states, the factor $R$ in~(\ref{oo}) should be replaced by a different factor; see below.}
with
\be\label{g}
\gamma(x)\equiv{\Gamma(x)\over\Gamma(1-x)}
\ee

A few comments are in order:

\begin{enumerate}

\item
Equation~(\ref{oo}) has the same form as eq. (3.7) in~\cite{Asrat:2017tzd}, with $\pi D(h_{p^2})\to R(w,j;h_{p^2})$,
and with the $h_{p^2}$ given in eq.~(\ref{hp});
it is obtained following the discussion around eqs. (3.4)--(3.6) and section 5.1 in~\cite{Maldacena:2001km},
and applying it to the discussion around eqs. (3.4)--(3.7) in~\cite{Asrat:2017tzd}.

\item
Strictly speaking, the result~(\ref{oo}) with~(\ref{a}) and~(\ref{hp}) is derived using (5.15),(5.16)
of~\cite{Maldacena:2001km}, which was obtained for $j={1\over 2}+is$.
The radial momentum $s$ is real for delta-function normalizable operators.
In this case, we have a continuous representation at $w=0$, prior to spectral flow in the underlying $SL(2,R)$
of the null quotient of $R^{1,1}\times AdS_3(\equiv{\cal M}_3)$,~(\ref{rads})--(\ref{m3}).
Non-normalizable operators can be obtained by continuation to real positive $is$, as in e.g.~\cite{Balthazar:2021xeh}.

\item
For delta-function normalizable operators, $j={1\over 2}+is$ and  $R(w,j;h_{p^2})$ is a phase:
$R(w,s;h_{p^2})=e^{i\delta(w,s;h_{p^2})}$(=(3.6) in~\cite{Maldacena:2001km} with $J=\bar J\to h_{p^2}$);
this is the phase shift in the scattering of long strings in target space,~${\cal M}_3$.
For non-normalizabe operators (${1\over 2}<j\in R$),
the discussion on p.~7,8 of~\cite{Asrat:2017tzd}, concerning the physics of such factors, is relevant here as well.

\item
For discrete states, $R$ in~(\ref{oo}) is not given by~(\ref{a}).
Operators that amount to discrete states in the null quotient of $R^{1,1}\times AdS_3(\equiv{\cal M}_3)$,~(\ref{rads})--(\ref{m3}),
are obtained from  normalizable operators in discrete representations at $w=0$,
a.k.a. with $m=\bar m=j+n$, $n=0,1,2,...$, ${1\over 2}<j<{k+1\over 2}$, in the underlying $SL(2,R)$, prior to spectral flow.
It thus follows from (5.18) in~\cite{Maldacena:2001km}~\footnote{With $J=\bar J\to h_{p^2}$ and $k-2\to k$ in appropriate places,
as required when going from the bosonic string in~\cite{Maldacena:2001km} to the superstring, considered here.}
that in this case, one should replace $R$ in~(\ref{oo}) by~\footnote{A.k.a. by the residues of poles from one of the gamma functions in~(\ref{a}), which amount to LSZ poles in the LST dual to the superstring theory on the asymptotically linear-dilaton spacetime 
${\cal M}_3$,~(\ref{rads})--(\ref{m3}); see~\cite{Aharony:2004xn} and references therein.}
\be\label{ar}
R\to\pi(2j-1+kw)B(j)\left({\Gamma(2j+n)\over n!\Gamma(2j)}\right)^2~,\qquad j=h_{p^2}-{k\over 2}w-n
\ee
where
\be\label{b}
B(j)={k\over\pi}{\nu^{1-2j}\over\gamma\left(2j-1\over k\right)}
\ee

\item
Recall that in the $w=0$ sector,~\cite{Asrat:2017tzd},
\be\label{2pf0}
\langle \hat O^{w=0,j=h_{p^2}}(p)\hat O^{w=0,j=h_{p^2}}(-p)\rangle_{target}=\pi D(h_{p^2})\gamma(1-2h_{p^2})\left(p^2\over 4\right)^{2h_{p^2}-1}
\ee
with the $h_{p^2}$ given by~(\ref{hp0}), and where
\be\label{d}
D(j)=(2j-1)B(j)
\ee
with the $B(j)$ in~(\ref{b}).
Indeed,~(\ref{ar}) collapses to $\pi D(h_{p^2})$ when $w=0$ (in which case one should also set $n=0$ in eq.~(\ref{ar})).

\item
Equation~(\ref{oo}), with~(\ref{a}) and~(\ref{hp}), is (4.15) in~\cite{Cui:2023jrb}, up to the factor of the
reflection coefficient, $R(w,j;h_{p^2})$, which is absent in~\cite{Cui:2023jrb},
and is in harmony with the dependence of the two-point correlator on the momentum at large momentum in eq.~(7) of~\cite{Aharony:2023dod},
which was obtained  using the JT-gravity formulation of $T\bar T$-deformed CFTs.

\end{enumerate}

To recapitulate, in the $w=1$ sector, the factor
\be\label{factor}
\gamma(1-2h-\lambda p^2)\left(p^2\over 4\right)^{2h-1+\lambda p^2}
\ee
in the 2pf
(see~(\ref{oo}), with~(\ref{a}) and~(\ref{hp}), for delta-normalizable and non-normalizable operators,
and with~(\ref{ar}) and~(\ref{hp}), for discrete states)
is the same factor that appears in the 2pf of $\lambda T\bar T$ deformed $CFT_2$,~\cite{Cui:2023jrb,Aharony:2023dod}.

For $w>1$, $\lambda\to\lambda/w$ in~(\ref{factor}), 
which is in harmony with a symmetric product of $\lambda T\bar T$ deformed $CFT_2$
(see e.g. the discussion around eqs. (6.5) and (6.6) in~\cite{Giveon:2019fgr} and references therein).

String theory also determines the prefactor of~(\ref{factor}).
For delta-normalizable states, it is a phase -- the phase shift in the scattering of long strings in string theory
on the null gauged~$R^{1,1}\times AdS_3(\equiv{\cal M}_3)\times{\cal N}$,~(\ref{rads})--(\ref{m3}).
For non-normalizable operators, it is the continuation of the bulk reflection coefficient,~(\ref{a}),
to imaginary radial momentum in target space, ${\cal M}_3$.
The poles from one of the gamma functions in the numerator of~(\ref{a}) signal the existence of normalizable states in the theory,
whose on-shell two point functions are~(\ref{oo}) with~(\ref{ar}) and~(\ref{b}).~\footnote{Recall,~\cite{Chakraborty:2023mzc}
and references therein, that the discrete states are known not to fit into a symmetric product.
However, these states have measure zero in the string spectrum, so one can say that the symmetric product describes generic states 
in perturbative string theory on ${\cal M}_3$}

On the other hand, the 2pf in the $w=0$ sector (see~(\ref{2pf0}) with~(\ref{hp0}) and~(\ref{d}) with~(\ref{b})) is
different,~\cite{Asrat:2017tzd}. In particular, it has a factor
\be\label{factor0}
\gamma\left(-\sqrt{(2h+1)^2+2\lambda kp^2}\right)\left(p^2\over 4\right)^{\sqrt{(2h+1)^2+2\lambda kp^2}}
\ee
instead of~(\ref{factor}), and its prefactor, $\pi D(h_{p^2})$,~(\ref{d}) with~(\ref{b}),
is the stringy contribution (a.k.a. non-perturbative in $\alpha'$)
to the (continuation of the) reflection coefficient.

The above is in harmony with previous observations
(see e.g.~\cite{Giveon:2017nie,Giveon:2019fgr,Apolo:2019zai,Cui:2023jrb,Chakraborty:2023mzc,Chakraborty:2023zdd}
and references therein and thereof),
and may shed more light regarding intriguing open problems in single-trace $T\bar T$ holography and their implications.

\vspace{10mm}

\section*{Acknowledgments}
This work was supported in part by the ISF (grant number 256/22), the BSF (grant number 2018068)
and the ISF center for excellence (grant number 2289/18).

\vspace{10mm}


\end{document}